\documentclass[10pt, conference, compsocconf]{IEEEtran}
\IEEEoverridecommandlockouts
 \usepackage{lineno}
\usepackage{float}
\usepackage{multicol}
\usepackage{graphicx}
\usepackage{amsmath,amssymb,amsfonts}
\usepackage{multirow}
\usepackage{amssymb}

\usepackage{textcomp}
\usepackage{lineno,hyperref}
\usepackage{color}

\ifCLASSINFOpdf
\else
\fi

\usepackage{stfloats}

\begin{document}
%
\title{HEMELB ACCELERATION AND VISUALIZATION FOR CEREBRAL ANEURYSMS}


\author{\IEEEauthorblockN{Sahar Soheilian Esfahani$^{\star}$, Xiaojun Zhai$^{\dagger}$, Minsi Chen$^{\ddagger}$, Abbes Amira$^{\star}$, Faycal Bensaali$^{\star}$ \\ Julien AbiNahed$^{\mathsection}$, Sarada Dakua$^{\mathsection}$, Georges Younes$^{\mathsection}$,  Robin A. Richardson$^{\|}$, Peter V. Coveney$^{\|}$\thanks{This paper was made possible by National Priorities Research Program (NPRP) grant No. 5-792-2-328 from the Qatar National Research Fund (a member of Qatar Foundation). The statements made herein are solely the responsibility of the authors.}}\\
\IEEEauthorblockA{\textit{$^{\star}$ College of Engineering, Qatar University, Doha, Qatar}}
\IEEEauthorblockA{\textit{$^{\dagger}$ School of CS and EE, University of Essex, Colchester, UK}}
\IEEEauthorblockA{\textit{$^{\ddagger}$ Department of Computer Science, University of Huddersfield, Huddersfield, UK}}
\IEEEauthorblockA{\textit{$^{\mathsection}$ Department of Surgery, Hamad Medical Corporation, Doha, Qatar}}
\IEEEauthorblockA{\textit{$^{\|}$ Centre for Computational Science, University College London, London, UK}}}


%


\maketitle

\begin{abstract}
A weakness in the wall of a cerebral artery causing a dilation or ballooning of the blood vessel is known as a cerebral aneurysm. Optimal treatment requires fast and accurate diagnosis of the aneurysm. HemeLB is a fluid dynamics solver for complex geometries developed to provide neurosurgeons with information related to the flow of blood in and around aneurysms. On a cost efficient platform, HemeLB could be employed in hospitals to provide surgeons with the simulation results in real-time. In this work, we developed an improved version of HemeLB for GPU implementation and result visualization. A visualization platform for smooth interaction with end users is also presented. Finally, a comprehensive evaluation of this implementation is reported. The results demonstrate that the proposed implementation achieves a maximum performance of 15,168,964 site updates per second, and is capable of speeding up HemeLB for deployment in hospitals and clinical investigations.

\end{abstract}
\section{Introduction}
\label{sec:intro}
A cerebral aneurysm is an abnormal focal dilation of a brain artery caused by a weakness in the blood vessel wall \cite{higashida2003you}. The number of patients who suffer from cerebrovascular disorders like cerebral aneurysms is growing in non-developed countries \cite{benkirane2015stroke}. A quick review of the statistics shows that people who already have or will develop brain aneurysm are between 1.5\% to 5\% of the general population \cite{mashiko2015development}. Another study shows that about 5\% of patients with growing cerebral aneurysm and 0.5-1.1\% of those with non-growing brain aneurysms will suffer from rupturing \cite{Statistics2018url}. 

An effective treatment of aneurysms are endovascular approaches which reduce the operative risks, pains and cost of hospitalization \cite{molyneux2002international}. These methods, which take advantage of intra-aneurysmal coils, may fail due to a lack of information about the aneurysm status. In recent years, the combination of coils with stents has been widely used as an efficient treatment which reorganizes the blood flow in and around the aneurysm \cite{kim2011stent}. In order to get the best treatment, the interventional radiologist has to identify the vascular geometry and estimate cerebral blood flow behavior such as pressure and velocity from 2D and/or 3D images. Presently, there are few approaches to measure these flows and related data intraoperatively. Consequently, diagnosis and treatment of aneurysms is highly dependent on the experience and skill of the radiologist.
\parskip 0pt

Modeling and simulation of fluid flow and hemodynamics can provide clinicians with more precise analysis and thus diagnosis of aneurysm effects. Simulation of fluid flow in large scale and complex geometries requires a physical model and substantial computing resources as well as efficient software. Recently, the lattice-Boltzmann method (LBM) is widely used for simulation of fluid flows. The LBM parallelizes efficiently making it useful for time-dependent simulation of large systems\cite{mazzeo2008hemelb}. In \cite{ZyncSoC:inpress}, the efficient hardware architecture of the LBM is introduced. HemeLB (Hemodynamic lattice-Boltzmann) is a massively parallel LBM fluid solver developed to simulate fluid flows in sparse and complex systems on large supercomputing resources \cite{mazzeo2008hemelb,groen2013analysing,patronis2018modelling}. The aim of designing HemeLB was to provide neurosurgeons with timely and clinically relevant assistance \cite{groen2018validation}. HemeLB blood flow simulations have been deployed on High Performance Computing (HPC) platforms including HECToR, Blue Waters, SuperMUC and ARCHER \cite{groen2013analysing,patronis2018modelling,groen2018validation}. It is required to optimize HemeLB on dedicated computational infrastructures in order to employ the software in clinical environments. Multicore platforms, General Purpose Graphical Processing Units (GPGPUs) and Field Programmable Gate Arrays (FPGAs) are probably today’s most powerful computational hardware found in various applications such as machine learning and artificial intelligence \cite{suchard2010understanding, zhai2018real, djelouat2018system}.

The objective of this paper is to develop a version of HemeLB optimized for GPU visualization and to evaluate it on a workstation with CUDA (Compute Unified Device Architecture) capable GPUs. To achieve this, additional functions and modifications have been made to the original HemeLB software to include a real-time GPU based rendering engine for running HemeLB in local environments. Additionally, a visualization platform intended to ease access to the simulation environment was also designed. This visualization platform will help clinicians to simply launch and use the HemeLB simulation software in hospitals. In order to evaluate the proposed implementation, a set of comprehensive tests using real patient data has been carried out on an Exxact Tensor workstation with four NVIDIA GPUs. 

The rest of the paper is organized as follows. In Section 2, the lattice-Boltzmann model used in HemeLB is briefly described. Section 3 presents the visualization in HemeLB. The experimental results and their evaluations are reported in Section 4 followed by conclusions in Section 5.
\section{HEMELB MODEL}
\label{sec:format}
\begin{figure}[htb]
 \centering
 \includegraphics[width=3.0cm,height=3.0cm]{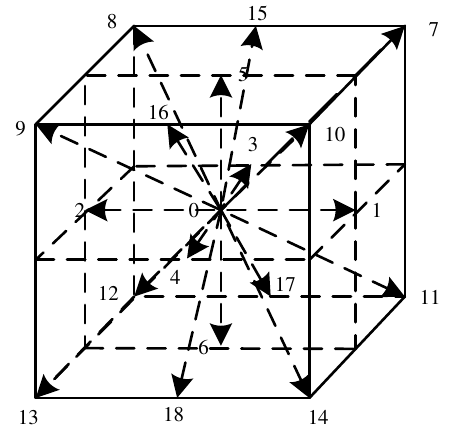}
 \caption{Lattice node of D3Q19 model \cite{mazzeo2008hemelb}.}
  \label{D3Q19Fig} 
\end{figure}

The LBM is a fast and efficient technique for the simulation of large and complex fluid systems, particularly via parallel implementations \cite{mazzeo2008hemelb}. HemeLB applies a highly parallelized implementation of the LBM, with performance optimizations for the sparse geometries common in vascular systems. The LBM represents the geometry as a lattice of fluid sites, each equipped with a single-particle distribution describing the propagation to neighbouring (or nearby) sites along a finite number of discrete lattice vectors. By modelling the evolution of these particle distributions over consecutive propagation and collision steps, we may recover the hydrodynamic behaviour expected for a continuum fluid. In the proposed system, we use a three dimensional lattice with 19 discrete velocities (D3Q19). Figure \ref{D3Q19Fig} presents an illustration of such a D3Q19 lattice.

HemeLB is implemented with different Boundary Conditions (BC) \cite{mazzeo2008hemelb}, such as Ladd iolets for velocity inlet BC \cite{ladd1994numerical} and Bouzidi-Firdaouss-Lallemand (BFL) for the interpolated wall collision BC \cite{bouzidi2001momentum}.
By the use of topology-aware two-level domain decomposition, HemeLB provides a good workload distribution for parallel implementation. In addition, the improvements made in HemeLB decrease redundant operations, improve pattern regularity and enhance intra-machine communications \cite{groen2013analysing}. 
\section{HEMELB IMPLEMENTATION}
\label{sec:majhead}
The benefits of GPU-oriented solutions such as processing speed and analysis of large datasets are achieved by the use of thousands of processor cores on a single chip \cite{suchard2010understanding}. In the following section, the real-time visualization of HemeLB on an Exxact Tensor workstation with four CUDA capable GPUs is discussed. 
\subsection{Architecture of Exxact Tensor workstation}
\begin{figure}
 \centering
 \includegraphics[width=4.2cm,height=4.2cm]{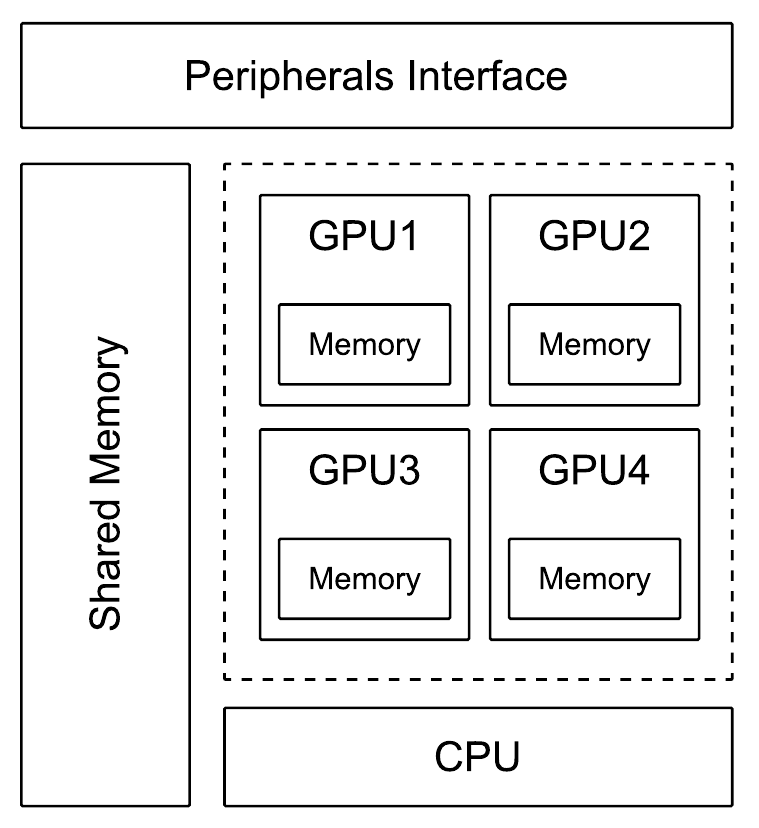}
 \caption{Simplified architecture of Exxact Tensor TWS-289059-DPN workstation.}
  \label{WorkstationSchematicFig} 
\end{figure}

The proposed work is implemented on an Exxact Tensor TWS-289059-DPN workstation which is known as a deep learning NVIDIA GPU solution \cite{ExxactTensorUrl}. The platform consists of one Intel core i7-5960X processor, 4 TB HDD, 1 TB SSD as well as four GeForce GTX 1080 Ti GPUs. Each NVIDIA GPU has 11 GB GDDR5X random access memory with 484 GB/s memory bandwidth and 3584 CUDA cores. The GPU base clock speed is 1481 MHz and the memory is running at 1376 MHz. Figure \ref{WorkstationSchematicFig} shows the simplified architecture of the workstation with four NVIDIA GPUs.
\subsection{HemeLB configuration}
The original version of HemeLB code is available in \cite{UCLHemelbRepository}. The \textit{openmpi}\ package and CUDA platform are required for the implementation of the proposed system. Once the package is built, compilation flags 
are needed to be set as boundary conditions. The boundary settings used in this experiment were \textit{INLET\_BOUNDARY}, \textit{WALL\_INLET\_BOUNDARY}, \textit{WALL\_BOUNDARY}, and \textit{USE\_VELOCITY\_WEIGHTS\_FILE} which were set to \textit{LADDIOLET}, \textit{LADDIOLETBFL}, \textit{BFL}, and \textit{ON} respectively. Table \ref{OtherSettingTable} represents the simulation, geometry and inlets settings used in the HemeLB. 
\begin{table}[htb]
\centering
\renewcommand{\arraystretch}{1.1}
\caption{HemeLB simulation, geometry and inlets settings.}
\label{OtherSettingTable}
\small
\begin{tabular}{cc}
 
\hline 
HemeLB settings & Values \\ 
\hline
Step length & $2\times10^{-5}$  s \\ 
Voxel size  & $100\times10^{-6}$  \\ 
Inlet velocity & file (see Figure \ref{BloodVelocityFig}) \\ 
\hline 
\end{tabular} 
\end{table}
\subsection{HemeLB visualization}
\subsubsection{HemeLB interface}
\begin{figure}[htb]
 \centering
 \includegraphics[width=6.1cm,keepaspectratio]{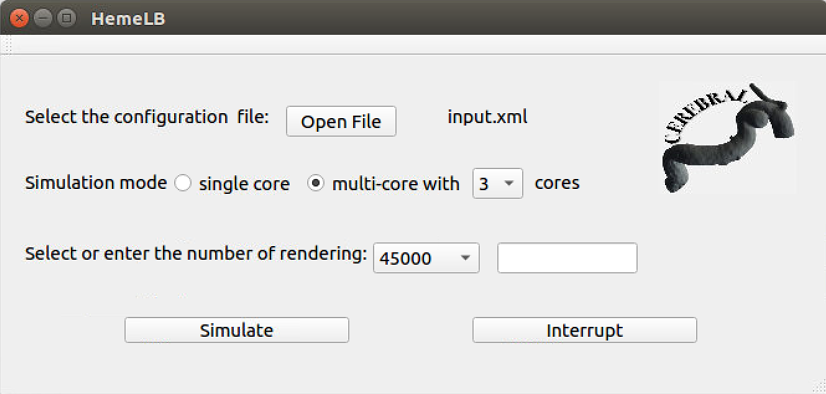}
 \caption{Visualization platform designed for real-time HemeLB results visualization.}
  \label{UIFig} 
\end{figure}

The original HemeLB code exposes a low-level technical interface useful for advanced users and accessed via the commandline. In order to enable clinicians to run HemeLB conveniently, a visualization platform has been developed. It provides end users with a smooth interaction environment for running the software. It allows clinicians and neurosurgeons to steer the visualization and manage the simulation operations. Figure \ref{UIFig} illustrates this visualization platform designed with the Qt cross-platform framework.

\begin{figure}[htb]
 \centering
 \includegraphics[width=7cm,height=4.5cm,keepaspectratio]{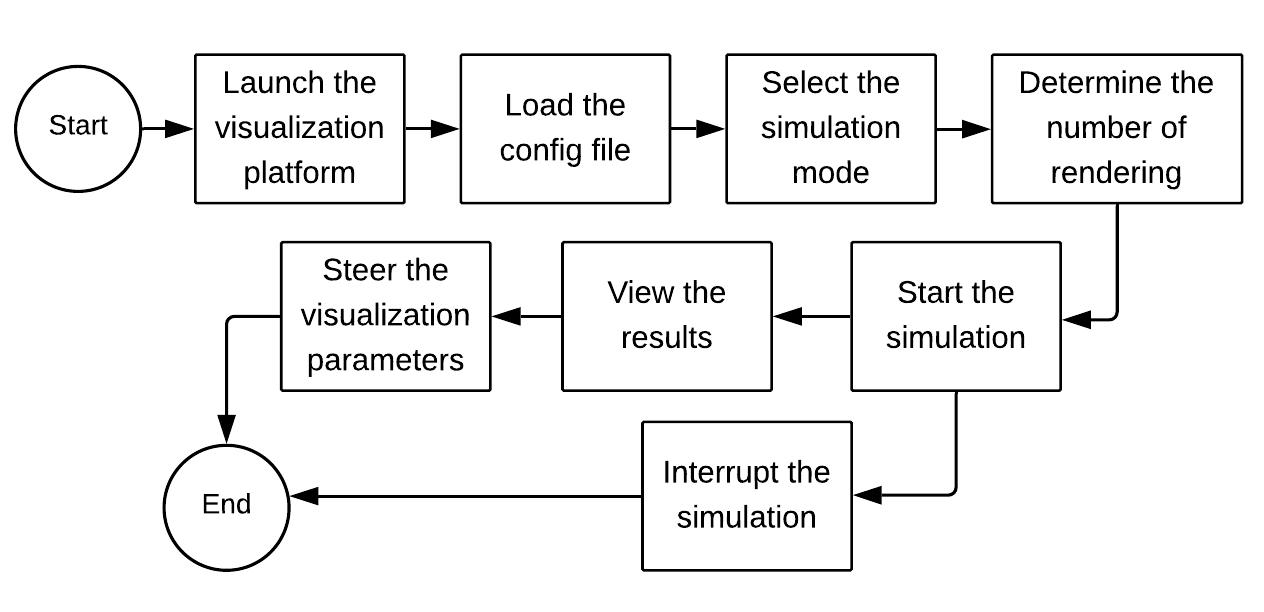}
 \caption{Flowchart of the HemeLB visualization platform.}
  \label{GUIFlowChartFig} 
\end{figure}

The visualization platform helps clinicians and neurosurgeons to launch HemeLB with a simple click. First, the user should upload the configuration file of the subject under examination, then set some input settings such as simulation mode and rendering number. This platform gives the control of the simulation mode to the clinicians. The user can choose a single core or multicore simulation mode before running the software. The number of cores in multicore mode can also be selected by the end user. Another option is the number of rendered frames which can be chosen from a list or entered by the clinician. Finally, the \textit{simulate}\ and \textit{interrupt}\ buttons give the control of simulation operation to the user. A flowchart of the designed platform is depicted in Figure \ref{GUIFlowChartFig}.
\subsubsection{Real time visualization}
\begin{figure}[htb]
 \centering
 \includegraphics[width=5.5cm,height=4.0cm]{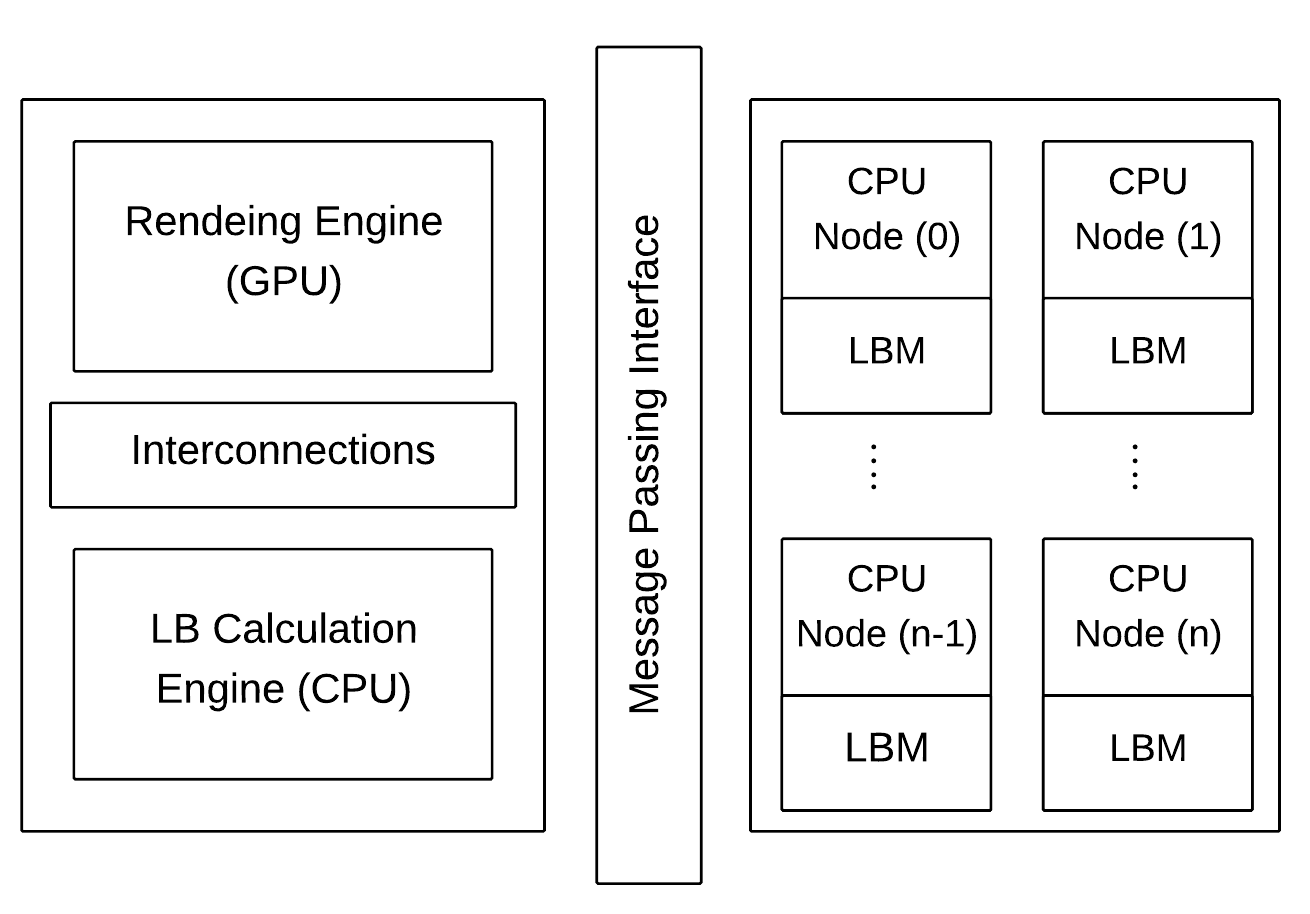}
 \caption{Architecture of proposed visualization framework.}
  \label{VisArchitectureFig} 
\end{figure}

Visualization of HemeLB is done on devoted CUDA capable GPUs. The LBM nodes and the visualization client communicate via the existing message passing interface in HemeLB. 
Figure \ref{VisArchitectureFig} shows the architecture of the proposed framework. 
In Figure \ref{VisArchitectureFig}, 
Lattice properties are computed and cached on each CPU node then communicated to $Node0$. This node schedules lattice data transfer from the compute nodes and handles the view steering. The incoming lattice data is shared with GPU nodes and rendering is performed at the same time.

\begin{figure}[htb]
 \centering
 \includegraphics[width=5.6cm,height=4.6cm]{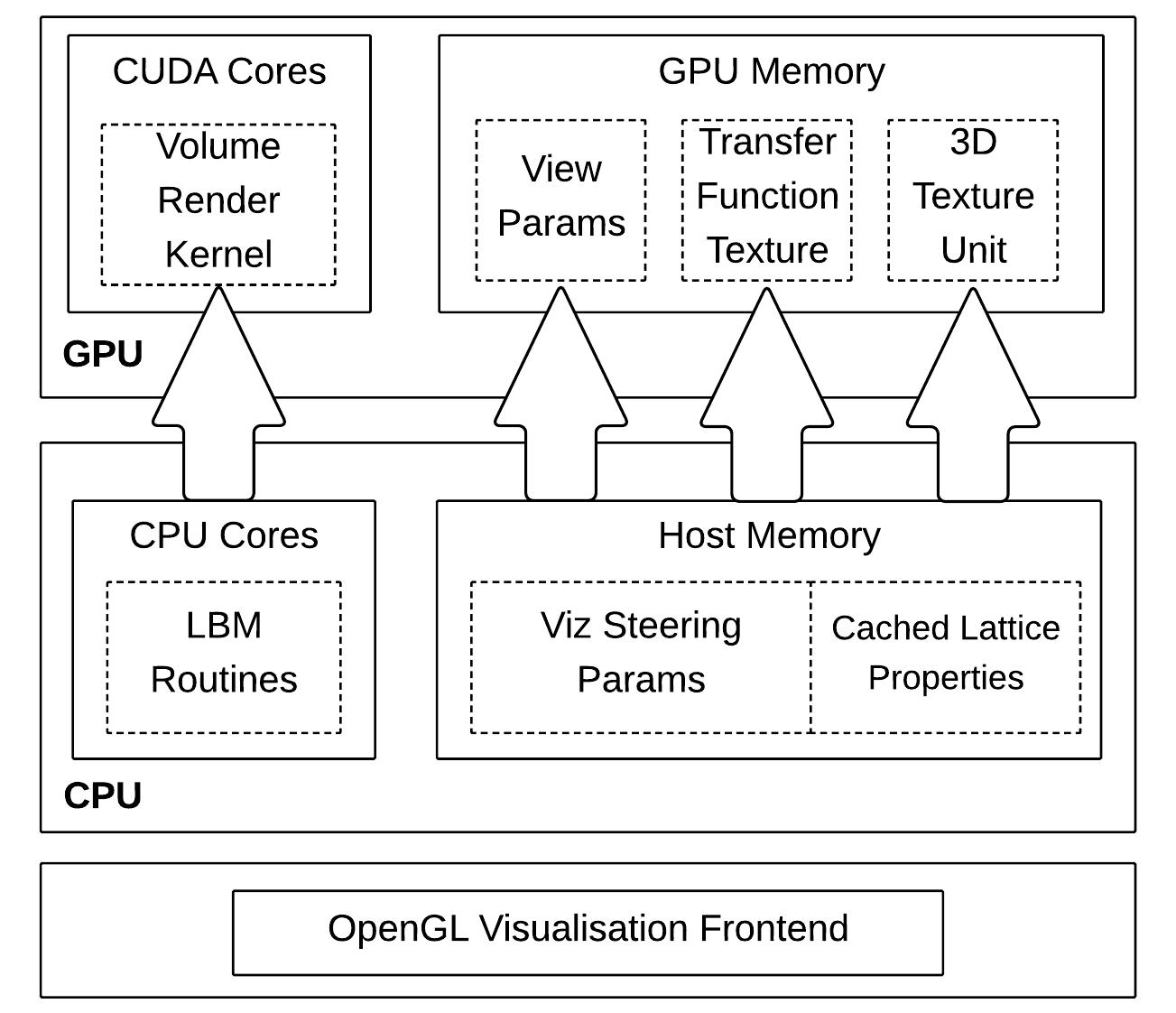}
 \caption{The three-tier internal architecture of visualization client.}
  \label{3tierArchFig} 
\end{figure}

The three-tier architecture of the visualization client is presented in Figure \ref{3tierArchFig}. The first tier is the OpenGL application and the middle one is the host layer which stores the application data such as steering parameters and cached lattice properties. Eventually, all the voxels for direct volume rendering are stored in the rendering layer which is depicted as the top tier. The CUDA cores in the rendering layer perform the volume rendering kernel based on the ray marching algorithm \cite{zhou2008real}. 

It is well known that the number of memory transfers is critical in real-time applications. In HemeLB visualization, the memory transfers include the transmission of lattice data to the visualization volume and a number of steering simulation. In order to optimize the memory transfers in the proposed framework, the following two-level memory access solution is deployed. In the first place, the visualization volume is stored in a 3D texture unit to enable the fast sampling of voxel values. 
Secondly, in order to store the viewing parameters and the look-up table, the GPU constant memory buffers are used. This allows Direct Memory Access (DMA) whose performance is comparable to reading from registers. 
The steerable parameters introduced in the visualized HemeLB include zooming, model rotation and adjustment of scaling and offset of the transfer functions. 
\section{EXPERIMENTAL RESULTS AND DISCUSSION}
\begin{figure}[htb]
\begin{minipage}[b]{.48\linewidth}
  \centering
  \centerline{\includegraphics[width=3.5cm,height=3.5cm]{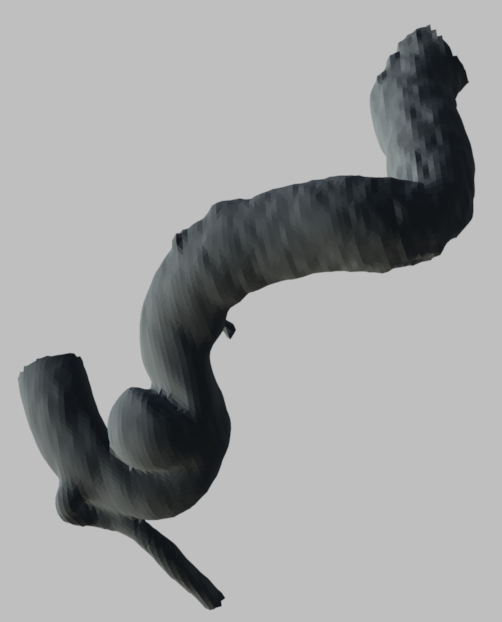}}
  \centerline{(a) Subject 16}\medskip
\end{minipage}
\begin{minipage}[b]{.48\linewidth}
  \centering
  \centerline{\includegraphics[width=3.5cm,height=3.5cm]{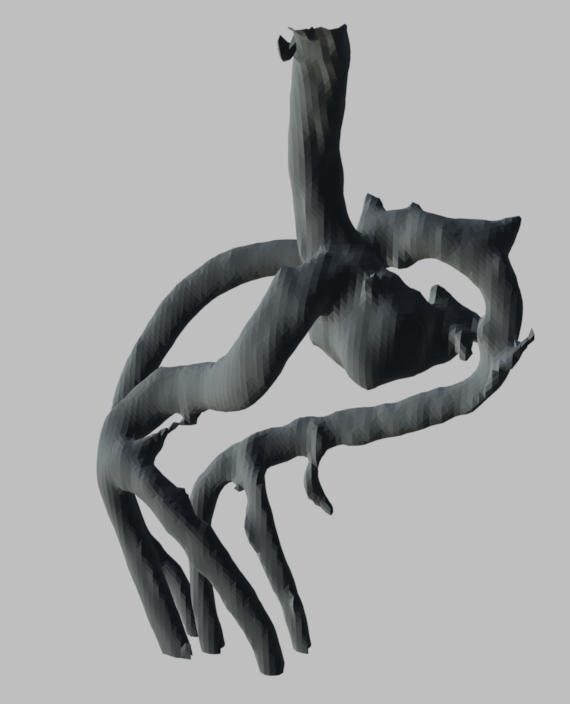}}
  \centerline{(b) Subject 23}\medskip
\end{minipage}

\caption{STL files used for evaluation of visualized HemeLB.}
\label{STLFig}
\end{figure}

\begin{figure}[htb]
 \centering
 \includegraphics[width=6.0cm, height=3.0cm]{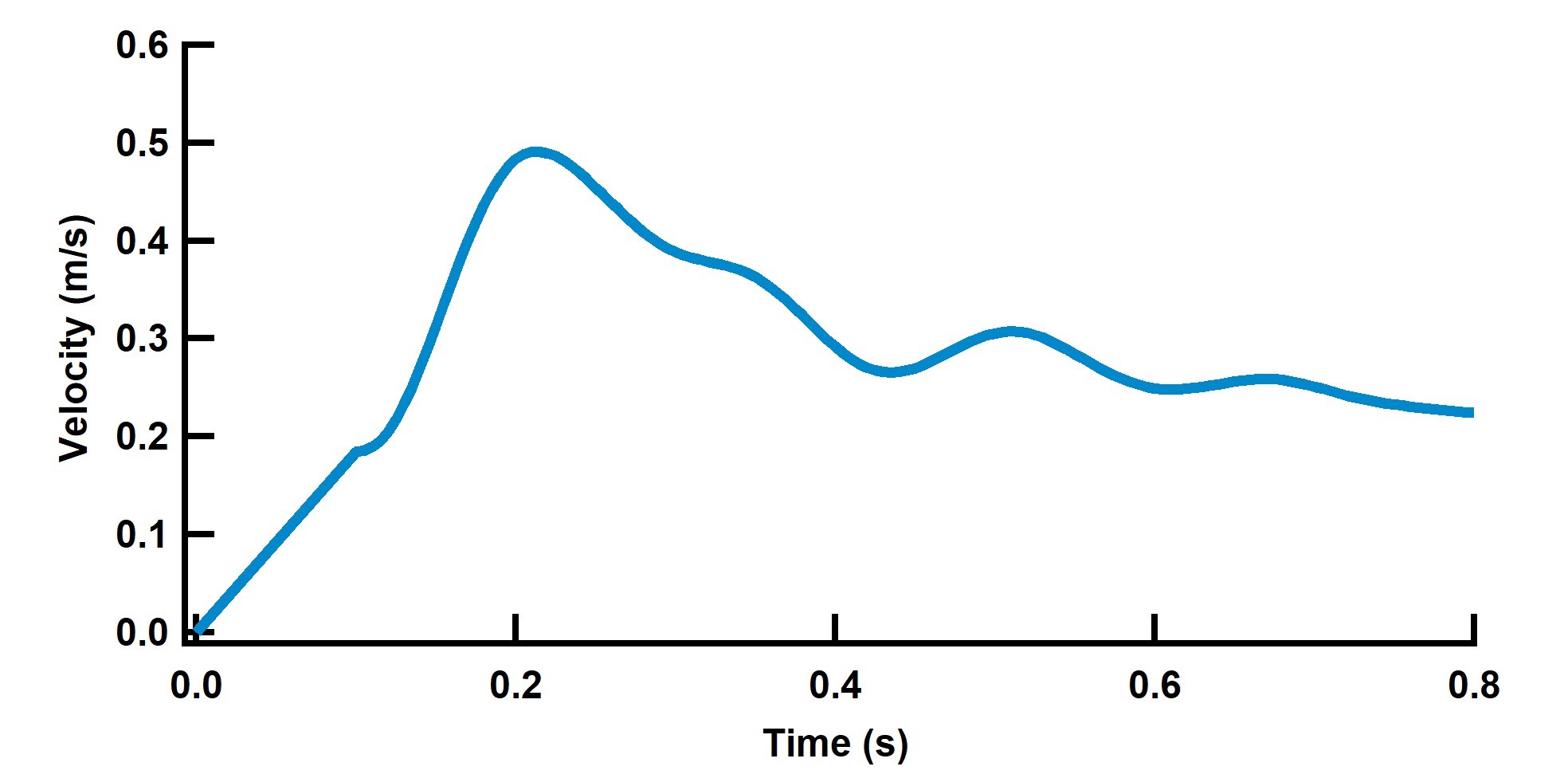}
 \caption{The inlet blood velocity applied in each patient test case.}
  \label{BloodVelocityFig} 
\end{figure}

Five subjects of 3D Rotational Angiography (3DRA) selected by Hamad Medical Cooperation (HMC) clinicians were used to evaluate the proposed visualization solution. Figure \ref{STLFig} represents two STereoLithograph (STL) files used to evaluate HemeLB visualization. In Table \ref{SimDomainTable} the number of lattice sites and blocks for each subject are presented. HemeLB was employed using a D3Q19 lattice model with simple bounce-back boundary condition and a fixed physical viscosity of 0.004 Pa.s. The inlet velocity applied for simulation with each STL input is illustrated in Figure \ref{BloodVelocityFig}. 
\begin{table}[htb]
\centering
\renewcommand{\arraystretch}{1.2}
\caption{Overview of simulation domains.}

\label{SimDomainTable}
\small
\begin{tabular}{ccc}
 
\hline 
Name & Number of lattice sites & Number of blocks \\ 
\hline
Subject 16 & 296,814 & 15,000 \\ 
Subject 23 & 132,910 & 5,320 \\ 

\hline 
\end{tabular} 
\end{table}

\begin{figure}
\begin{minipage}[b]{.48\linewidth}
  \centering
  \centerline{\includegraphics[width=3.8cm,height=3.8cm]{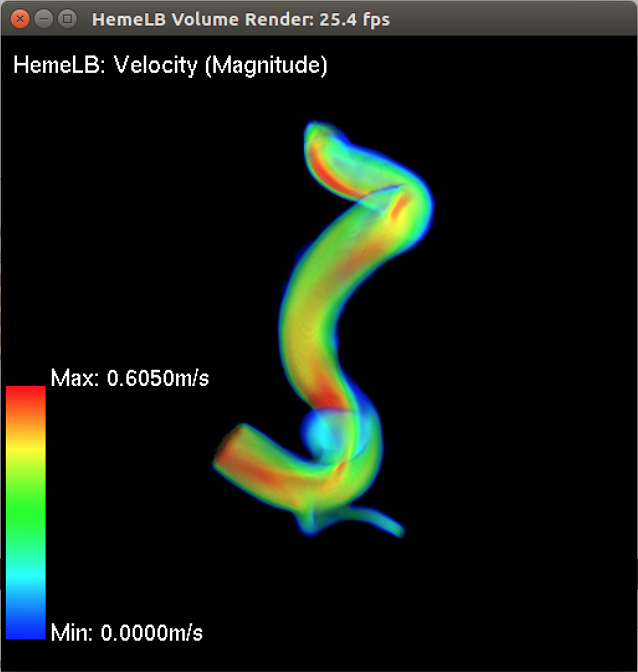}}
  \centerline{(a) Subject 16}\medskip
\end{minipage}
\begin{minipage}[b]{.48\linewidth}
  \centering
  \centerline{\includegraphics[width=3.8cm,height=3.8cm]{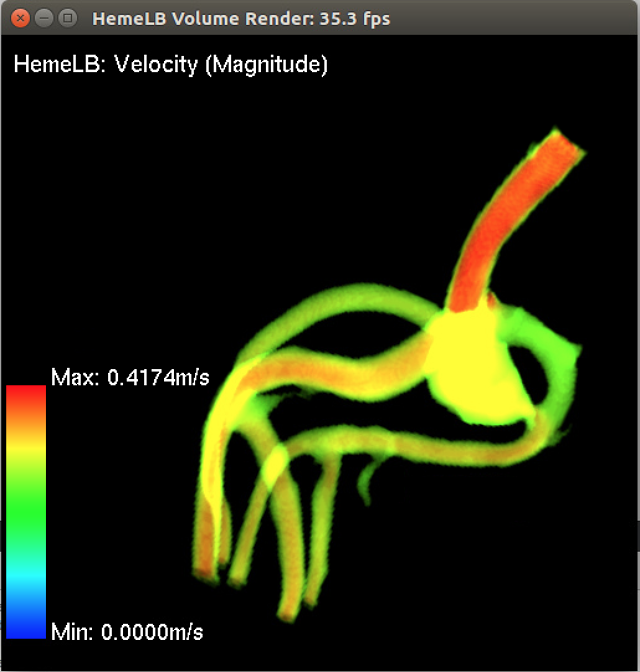}}
  \centerline{(b) Subject 23}\medskip
\end{minipage}
\caption{Results of HemeLB visualization framework.}
\label{ResFig}
\end{figure}

Table \ref{TimingDataTable} shows the performance for two subjects simulated on the Exxact Tensor workstation. The maximum performance of 15,168,964 Site Updates Per Second (SUPS) is achieved in this evaluation. As it is depicted in Table \ref{TimingDataTable}, the larger geometry has the highest SUPS, although the timing performance for different subjects is similar. The simulation results of two geometries obtained using the proposed version of HemeLB are illustrated in Figure \ref{ResFig}. 
\begin{table}[htb]
\centering
\renewcommand{\arraystretch}{1.1}
\caption{Simulation timing data.}
\label{TimingDataTable}
\small
\begin{tabular}{cccc}
 
\hline 
Name & Total time (s) & Number of steps & SUPS \\ 
\hline
Subject 16 & 906 & 46,302 & 15,168,964 \\ 
Subject 23 & 297 & 30,558 & 13,674,962\\ 

\hline 
\end{tabular} 
\end{table}
\section{CONCLUSION}
In this paper, a solution for designing and implementing a real-time visualization version of HemeLB on a workstation is presented. The proposed implementation enables HemeLB to be exploited locally in hospitals rather than in a distributed environment. In addition, a simple visualization platform for the HemeLB environment is also introduced in this work. This platform allows clinicians to launch the simulation, and to view and steer the visualized results in a user friendly and smooth manner. The reported results in this experience are obtained from tests performed on real patient data. The results achieve a maximum performance of 15,168,964 site updates per second and demonstrate that the proposed implementation is capable of supporting HemeLB in clinical environments. A working system which can be used in hospitals for training and practicing the cerebral surgeries can be developed as a future work. 
\label{sec:ref}
\bibliographystyle{IEEEbib}
\bibliography{strings,refs}

\end{document}